\documentclass{eptcs}

\usepackage{epsfig}
\usepackage{amsmath}
\usepackage{amssymb}
\usepackage{amscd}
\usepackage{galois}
\usepackage{graphics}
\usepackage{algorithmic,algorithm}
\usepackage{stmaryrd}

\newcounter{examplecount}
\newenvironment{example}{\smallskip\stepcounter{examplecount}\textbf{Example \arabic{examplecount}}}{\smallskip}

\newcounter{propertycount}
\newenvironment{property}{\smallskip\stepcounter{propertycount}\textbf{Property \arabic{propertycount}}}{\smallskip}

\newcommand{\code}[1]{\texttt{#1}}
\newcommand{\sem}[1]{\llbracket\texttt{#1}\rrbracket}
\newcommand{\asem}[1]{\llbracket\texttt{#1}\rrbracket^{a}}
\newcommand{\notOverflowModels}{\models_\mathit{notOverflow}}
\newcommand{\notOverflow}[2]{#1 \notOverflowModels #2}
\newcommand{\notnotOverflow}[2]{#1 \not\notOverflowModels #2}

\newcommand{\refFig}[1]{Fig.~\ref{fig:#1}}

\newcommand{\refSec}[1]{Sec.~\ref{sec:#1}}
\newcommand{\refsec}[1]{\refSec{#1}}

\newcommand{\refFormula}[1]{(\ref{formula:#1})}

\newcommand{\refEq}[1]{\refFormula{#1}}

\newcommand{\clousot}{\code{cccheck}}
\newcommand{\Clousot}{\clousot}
\newcommand{\astree}{\code{Astree}}
\newcommand{\Astree}{\astree}

\newcommand{\qed}{}

\title{Automatic Repair of Overflowing Expressions\\with Abstract Interpretation}

\author{Francesco Logozzo 
  \institute{Microsoft Research, Redmond, WA, USA}
\and 
Matthieu Martel
\institute{
 Universit\'e de Perpignan Via Domitia, DALI, Perpignan, France\\
 Universit\'e Montpellier II \& CNRS, LIRMM, UMR 5506, Montpellier, France}
}

\def\titlerunning{Automatic Repair of Overflowing Expressions with Abstract Interpretation}
\def\authorrunning{F. Logozzo \& M. Martel}

\begin{document}

\algsetup{indent=2em}
\pagestyle{plain}

\maketitle

\begin{abstract}
We consider the problem of synthesizing provably non-over\-flow\-ing \emph{integer} arithmetic expressions or Boolean relations among integer arithmetic expressions.
First we use a numerical abstract domain to infer numerical properties among program variables.
Then we check if those properties guarantee that a given expression does not overflow.
If this is not the case, we synthesize an equivalent, yet not-overflowing expression, or we report that such an expression does not exists.

The synthesis of a non-overflowing expression depends on three, orthogonal factors: the input expression (\emph{e.g.}, is it linear, polynomial, \dots ?), the output expression (\emph{e.g.}, are case splits allowed?), and the underlying numerical abstract domain -- the more precise the abstract domain is, the more correct expressions can be synthesized.

We consider three common cases: (i) linear expressions with integer coefficients and intervals; (ii) Boolean expressions of linear expressions; and (iii) linear expressions with templates.
In the first case we prove there exists a complete and polynomial algorithm to solve the problem.
In the second case, we have an incomplete yet polynomial algorithm, whereas in the third we have a complete yet worst-case exponential algorithm.
\end{abstract}

\section{Introduction}

Unwarranted integer overflows are a common source of bugs even for the most experienced programmers.
Programmers have the tendency of forgetting that \emph{machine} integers behave differently than \emph{mathematical} integers and that apparently innocuous expressions may lead to hard to debug bugs in the program.
Consider for instance the statement below, where \code{y} is a negative $32$-bits integer
\begin{equation}
  \code{x = -y;}
\label{formula:negation}
\end{equation}
One may then expect that \code{x} is always a positive value, and also that $\code{x} \neq \code{y}$.
However, this is false.
When \code{y} is the minimum value  $-2^{31}$, then $\code{x} = -2^{31} = \code{y}$, \emph{i.e.}, the result of the negation of a negative integer is negative!
This is not an artificial example.
The \code{Math.abs} function in the standard Java library implements the absolute function value function according to the  common \emph{mathematical} definition (if the input is non-negative return it, otherwise return its negation).
As a consequence, \code{Math.abs} may return a negative value, very likely breaking most callers relying on the (somehow obvious) fact that the absolute value is always non-negative.

The mismatch between the mathematical interpretation and the machine interpretation comes by the fact that the expression evaluation may originate in an \emph{arithmetic overflow}: the result of the expression may be too large (or too small) to be exactly represented on $b$ bits.

Most abstract interpretation-based static analysis tools like \Clousot~\cite{FahndrichLogozzo10} or \Astree~\cite{astree} can detect potential arithmetic overflows.
They first analyze the program using some numerical abstract domain (\emph{e.g.}, Intervals~\cite{CousotCousot77}, Pentagons~\cite{LogozzoFahndrich08-2}, Octagons~\cite{Mine-HOSC06}) to infer ranges and relations among program variables at each program point.
Then, they use such information to prove that the evaluation of an arithmetic expression may never result into an overflow.
If they cannot prove it, they emit a warning.
For instance, in the example above \clousot\ warns about the possible negation of the \code{MinValue}.

In this paper, we want to push it a step further. 
We envision static analysis tools not only  reporting possible arithmetic overflows, but \emph{also} suggesting  fixes for them.
The suggested fixes are \emph{verified repairs} in the sense of~\cite{LogozzoBall-OOPSLA12}.
A verified repair is an expression which is equivalent to the original one when interpreted over the mathematical integers $\mathbb{Z}$, but which does not overflow when evaluated according to the given programming language semantics.

In general, the repair depends on three orthogonal factors: 
\begin{itemize}
\item[(i)] the input expression language, $\mathcal{I}$,
\item[(ii)] the output expression language, $\mathcal{O}$,
\item[(iii)] the available semantic information, $\mathcal{S}$, \emph{i.e.}, the underlying abstract domain and the abstract state inferred by the analyzer.
\end{itemize}
For instance, let us consider the repairing of~\refFormula{negation} where the input and output expression language are arithmetic expressions with \emph{only} the $4$ operations.
The semantic information is 
\[
\mathcal{S} = \{\code{y} \in [-2^{31}, -1] \}.
\]
Under these conditions, there is no way to repair the expression.
Our algorithm in~\refsec{sum} will prove that there is no way to fix the expression~\refFormula{negation}.
Nevertheless, if we change the hypotheses, allowing the  output language $\mathcal{O}$ to include expression casting, then  $\code{-((long) y)}$ is a verified repair for $\code{-y}$ under $\mathcal{S}$.
The arithmetic overflow disappears as $2^{31}$ is exactly representable in $64$ bits\footnote{Note that this repair may require changing the nominal type of \code{x} if it is not defined as a \code{long}.}, and the semantics coincides with that over $\mathbb{Z}$. 
Alternatively, we can imagine adding the conditional expression to $\mathcal{O}$.
Then 
\[
\code{y} == \code{MinValue} ?\ \code{0} :\ -\code{y}
\]
 is a repair in that the expression is guaranteed to not overflow, but it is \emph{not} a verified repair as the $\mathbb{Z}$ interpretation and the machine one disagree.

\begin{example}
\label{ex:Validation}
Let us consider the code in~\refFig{Validation}, returning a sub-array of \code{arr} made up of \code{count} elements from index \code{start}. 
The careful programmer added preconditions (using CodeContracts,~\cite{BarnettFahndrichLogozzo-SAC10}) to protect its code against buffer overruns: the starting index and the count should be non-negative (to avoid buffer underflows) and the subsegment to extract should be included in the original array (to avoid buffer overflows).
However, when \code{start} and \code{count} are very large,  \code{start + count} may result into an arithmetic overflow, \emph{i.e.}, it evaluates to a negative value.
As a consequence, the third precondition is trivially satisfied (an array length is always non-negative), but the program will probably go wrong,  because of some buffer underflow later in the execution.
 \qed
\end{example}

Again, sound static analysis tools like~\Clousot\ will spot the possible arithmetic overflow in the example above.
Our goal here is to synthesize an arithmetic expression matching the intent of the programmer (\emph{i.e.}, the semantics over $\mathbb{Z}$).
The algorithm we introduce in~\refsec{Relations} will synthesize the expression:
\begin{equation}
  \code{start <= arr.Length - count}
  \label{formula:ValidationFix}
\end{equation}
which is guaranteed to be arithmetic overflows-free.
The reason for that is that the abstract element
\[
\mathcal{S} = \{  \code{start} \in [0, 2^{31}-1], \code{count}  \in [0, 2^{31}-1], \code{arr.Length} \in [0, 2^{31}-1]\}
\]
 implies that 
\[
\code{arr.Length} -\code{count}
\]
 can never be too small to cause an arithmetic overflow.

In general, on machine arithmetics \refEq{ValidationFix} \emph{is not} equivalent to the third postcondition of~\refFig{Validation}.
Therefore, the repair cannot be done in a purely syntactic way, but it requires some semantic knowledge.

\begin{figure}[t]
\begin{verbatim}
int[] GetSubArray(int[] arr, int start, int count)
{
  Contract.Requires(0 <= start);
  Contract.Requires(0 <= count);
  Contract.Requires(start + count <= arr.Length);
  // ... rest of the code omitted ... 
}
\end{verbatim}
\caption{A parameter validation incorrect because of overflows.}
\label{fig:Validation}
\end{figure}

\begin{example}\label{ex3}
Let us consider $\mathcal{I} = \mathcal{O}$ to be the language of arithmetic additions.
For simplicity, let us assume to have  $4$ bits signed integers.
As a consequence $-8$ and $7$ are respectively the smallest and the largest representable natural numbers. 
The sum
\begin{equation}
  \code{x}_1 + \code{x}_2 + \code{x}_3 + \code{x}_4 + \code{x}_5+ \code{x}_6 + \code{x}_7
  \label{formula:sum}
\end{equation}
with the  semantic knowledge
\[
\mathcal{S} = \{  \code{x}_1 \in [-2,3],  \code{x}_2 \in[-1,0], \code{x}_3 \in[1,2], \code{x}_4 \in [-3,-1], 
 \code{x}_5 \in [-3, -2], \code{x}_6 \in[-1,1], \code{x}_7 \in [2,4]
\}
\]
may overflow when adding $\code{x}_6$ to the partial sum --- assuming left-to-right evaluation order.\qed
\end{example}

Our algorithm in~\refsec{sum} will automatically synthesize the following sum:
\begin{equation}
  \code{x}_3 + \code{x}_7 + \code{x}_4 + \code{x}_5 + \code{x}_2+ \code{x}_6 + \code{x}_1
  \label{formula:sumRepaired}
\end{equation}
which is a verified repair, as~\refEq{sum} and~\refEq{sumRepaired} coincide with the $\mathbb{Z}$ interpretation, but~\refEq{sumRepaired} does not overflow.

\paragraph{Main Results.}
First, we will define the problem of the automatic repairing of overflowing expressions starting from the semantic knowledge inferred by a static analyzer~(\refSec{problem}).
Next, we will focus our attention on the common case of linear expressions with integer coefficients.
We show that when the underlying abstract domain is Intervals~(\refSec{sum}) then there exists a \emph{complete and polynomial} algorithm to repair possibly overflowing expressions -- this result was quite surprising for us.
When the input language is widened to consider relations among those linear expressions, then the algorithm is still polynomial, but incomplete~(\refSec{Relations}).
On the other hand, when the underlying abstract domain is refined to Octagons, we have a complete yet worst-case exponential algorithm --- this result can be easily generalized to template-based numerical domains~(\refSec{Octagons}).
We will conclude evaluating a prototype implementation on a set of benchmarks~(\refSec{experiments}).

\section{Preliminaries}
\label{sec:notation}

\paragraph{Syntax.}
Without any loss of generality we assume a strongly typed while-language \code{W} with expressions \code{E}.
Expressions are either \emph{arithmetic} expressions (\emph{e.g.},  $5*\code{x} + 1$) or \emph{relational} expressions (\emph{e.g.}, $\code{x} + 2 < 3 * \code{y}$).
Variables are declared to belong to some integral type $M_{b}$.
Expressions are well typed, too. 

An integral type can be a signed or unsigned integer of $b$ bits. 
The smallest representable unsigned value over $b$ bits is $0$ and the largest is $2^{b}-1$.
The smallest representable signed is $-2^{b-1}$ and the largest one is $2^{b-1}-1$.
We denote by $\underline{M}_b$ and  $\overline{M}_b$  the smallest  and the largest values of $M_\code{b}$.
When clear from the context, we will omit the subscript $b$.

\paragraph{Semantics.}
A program state either denotes an error  or maps  variables to values, \emph{i.e.}, 
\[
\Sigma=  \{\code{err} \} \cup (\code{Vars} \rightarrow \mathbb{Z}).
\]
The only way to reach the error state \code{err} is by means of an arithmetic overflow --- for the sake of simplicity we treat division by zero as a particular case of overflow.
We assume a semantic evaluation function 
\[
\mathsf{eval} \in \code{E}\times\Sigma \rightarrow \mathbb{Z} \cup \{ \code{err} \}.
\] 
The function $\mathsf{eval}$ respects a fixed left-to-right evaluation of arithmetic expressions (like  C\# or Java, but unlike C).

If an arithmetic overflow is encountered during the evaluation of  $\code{e} \in \code{E}$ in $\sigma \in \Sigma$ then   $\mathsf{eval}(\code{e}, \sigma) =\code{err}$.

We define the relation $\notOverflowModels \in \wp(\wp(\Sigma) \times \code{E})$ as
\[
\notOverflow{S}{\code{e}} \Longleftrightarrow \forall \sigma \in S.\ \mathsf{eval}(\code{e}, \sigma) \neq \code{err}.
\]

The \emph{concrete} semantics of a program \code{P} associates each program point $\code{pc} \in \code{PP}$ with the set of reachable states $\subseteq \Sigma$.
Formally: $\sem{P} \in \code{PP} \rightarrow \wp(\Sigma)$.

An abstract domain $\langle \mathcal{A}, \sqsubseteq\rangle $ over-approximates sets of states.
In the abstract interpretation framework, this is formalized by a Galois connection, \emph{i.e.}, by two monotonic functions $\alpha, \gamma$ such that 
\[
  \forall S \in \wp(\Sigma).\ S \subseteq \gamma \circ \alpha(S)\ \text{and}\ \forall a \in \mathcal{A}.\ \alpha \circ \gamma(a) \sqsubseteq a.
\]
The \emph{abstract} semantics of \code{P} associates each program point with a \emph{sound} approximation of the reachable states.
Formally, $\asem{P} \in \code{PP} \rightarrow \mathcal{A}$ is such that
\[
\forall \code{pc} \in \code{PP}.\ \sem{P}(\code{pc}) \subseteq \gamma(\asem{P}(\code{pc})).
\]

\section{The Problem}
\label{sec:problem}

In order to define the problem of repairing (or synthesis) of non-overflowing expressions, we should make all the underlying assumptions  explicit.

First, we assume the program \code{P} is analyzed using  a sound static analyzer with an underlying abstract domain $\mathcal{A}$ \cite{CousotCousot77,Schmidt09}.

Second, we  denote the set of potential arithmetic overflows by $\mathbb{A} = \{ \langle\code{pp}, \code{e} \rangle \mid \code{pc} \in \code{PP} \}$.

Third,  we define $\mathbb{O}$, the subset of  $\mathbb{A}$ for which we cannot prove the absence of overflows, \emph{using} the abstract domain $\mathcal{A}$:
\[
\mathbb{O} = \{ \langle \code{pp}, \code{e} \rangle \in \mathbb{A} \notnotOverflow{\mid \gamma(\asem{P}(\code{pp}))}{\code{e}} \}.
\]

Usually, static analyzers stop here: 
They simply report the assertions $\mathbb{O}$ as possible overflows. 
We want to move it a step further.

We want to propose a verified repair for the expressions in $\mathbb{O}$.
In general, we cannot hope to repair arbitrary arithmetic expressions, so we focus our attention on a given set of input expressions $\mathcal{I}$.
Examples of $\mathcal{I}$  are sums with unary coefficients $\code{U} = \{ \code{e} \mid \code{e} = \sum_i \code{x}_i   \}$ and  linear arithmetic with integer coefficients $\code{L} = \{ \code{e} \mid \code{e} = \sum_i a_i \cdot \code{x}_i, a_i \in \mathbb{Z}   \}$.
The repaired, or non-overflowing output expression belongs to a set of expressions $\mathcal{O}$. 

We require that $\mathcal{I} \subseteq \mathcal{O} \subseteq \code{W}$, \emph{i.e.}, the output expressions are at least as expressive as the input ones, and both of them are expressible in our programming language.
We assume that the  expressions in $\mathcal{I}$ and $\mathcal{O}$ are side-effect free.

Our problem is to find for each expression \code{e} in $\mathbb{O}$ an equivalent expression in $\mathcal{O}$ which is provably non-overflowing according to the semantic information inferred by the static analyzer at program point \code{pp}.
Formally, we want an algorithmic characterization of the set of repairs:
\[
\mathbb{R}_{\langle{\mathcal{I}, \mathcal{O}, \mathcal{A}\rangle}} = \{ \langle \code{pp}, \code{e}' \rangle \mid \langle \code{pp}, \code{e} \rangle \in \mathbb{O}, \code{e} \in \mathcal{I}, \code{e}' \in \mathcal{O}, \code{e} \equiv_\mathbb{Z} \code{e}',  \notOverflow{\gamma(\asem{P}(\code{pp}))}{\code{e}'}    \}
\]
where $\equiv_\mathbb{Z}$ denotes the equivalence of expressions when they are interpreted over natural numbers and \emph{not} machine numbers.
In practice, we are interested in non-trivial solutions to $\mathbb{R}$, or in non-trivial under-approximations of $\mathbb{R}$.
An example of a trivial solution is  the brute-force generation of all the possible parsings $\code{e}'$ of an expression $\code{e} \in \mathcal{I}$ followed by testing  $\code{e}'$  whether or not may overflow.
This solution is exponential in the size of~\code{e}, a situation we want to avoid.
An example of a trivial under-approximation is the empty set --- \emph{i.e.}, no expression is repaired.

\section{Linear Expressions with Integer Coefficients}
\label{sec:sum}

We begin by focusing our attention  on the very common case when $\mathcal{I} = \mathcal{O} = \code{L}$, \emph{i.e.}, the input and the output languages are linear expression with  \emph{integer} coefficients.
We assume the underlying abstract domain $\mathcal{A}$ to be the abstract domain $\mathsf{Intv}$ of intervals~\cite{CousotCousot77}.
In this section we show that there exists a \emph{complete} and \emph{polynomial} algorithm for $\mathbb{R}_{\langle \code{L}, \code{L}, \mathsf{Intv}\rangle}$ -- Algorithm~\ref{algosum}.

The first step of Algorithm~\ref{algosum} is a pre-processing step to get rid of multiplications of an integer constant to a variable.
Given an expression $\code{e} \in \code{L}$, \emph{i.e.}, 
\[
\code{e}=\sum_{i=1}^n a_i \cdot \code{x}_i \qquad \text{ with } a_i\in \mathbb{Z} \text{ and } 1\le i\le n
\]
each term of the sum  $a_i\cdot \code{x}_i$ is replaced by the corresponding sum:
\[
a_i\cdot \code{x}_i=\underbrace{\code{x}_i+\ldots + \code{x}_i}_{a_i \ \text{times}}.
\]
Without loss of generality, we assume that all the $a_i$ are positive.
Otherwise: 
\begin{itemize}
\item[(i)] if $a_i$ is zero, then it is trivial to remove $\code{x}_i$,
\item[(ii)] if $a_i$ is negative, then we record it by negating the interval $\mathcal{S}(\code{x}_i)$.
\end{itemize}
The pre-processed expression is then in the form of $\sum_{i=1}^k \code{y}_i$ where $k=\sum_{i=1}^n a_i$ and $\code{y}_i$ are variables appearing in \code{e}.

After pre-processing, Algorithm \ref{algosum} partitions the (indexes of the) variables appearing in the expression into three sets: 
the positive variables $Y_{>0}$, the negative variables $Y_{<0}$,
and the variables  that can be positive, negative, or zero $Y_{\top}$. 
Let $\underline{x}$ and $\overline{x}$  denote respectively the lower and upper bounds of an interval $x$.
Then we can formally define the above sets as:
\[
Y_{> 0}=\{ i \mid 1\le i\le k,\ \underline{\code{y}_i}> 0\}, Y_{<0}=\{ i \mid 1\le i\le k,\ \overline{\code{y}_i}< 0\}, Y_{\top}=\{ i \mid 1\le i\le k,\ \underline{\code{y}_i}\le 0\ \wedge\ \overline{\code{y}_i}\ge 0\}.
\]
We use the shortcut $R=Y_{>0}\cup Y_{<0}$ for the set of variables that do contain zero.

The main loop of  Algorithm \ref{algosum} constructs a permutation of terms in the input sum such that the partial sum $s$ is kept as far as possible 
from $\underline{M}$ and $\overline{M}$.
Initially, the partial interval sum $s$ is set to $[0,0]$ and $\pi$ is set to the identity function. 
The algorithm iteratively selects an element $\code{y}_i$ with $i\in R$ such that $s+\code{y}_i\subseteq M$ (while loop of Line 5). 
The index $i$ is removed from $R$ and the value of $\code{y}_i$ is added to partial (interval) sum $s$.
The permutation $\pi$ records which term is selected at each iteration.
If, at some iteration, there exists no term $\code{y}_i$ with $i\in R$ such that $s+\code{y}_i\in M$ then the algorithm fails. We will see
in Property \ref{propsum} that, in this case, there exists no solution. 
Once all the elements with indexes in $R$ have been selected, the algorithm selects the elements whose indexes belong to $Y_{\top}$ in any order (for loop of Line 13). 
If the sum is not included in  $M$ then the algorithm fails.

\begin{algorithm}[tb]
\caption{Sum Refactoring}\label{algosum}
\begin{algorithmic}[1]
\REQUIRE An expression $\code{e} \in \code{L}$ and a map $\mathcal{S} \in \code{Vars} \rightarrow \mathbb{Z} \times \mathbb{Z}$
\smallskip

\ENSURE A permutation $\pi \in [1,k]\rightarrow [1,k]$ indicating in which order to add the elements $\code{y}_i$ to avoid overflows, or fail if it does not exist.
\medskip

\STATE Transform \code{e} into a sum such that $\code{e}=\sum_{i=1}^k \code{y}_i$
\STATE Create $Y_{> 0}$, $Y_{< 0}$ and $Y_{\top}$
\STATE $R \leftarrow Y_{> 0}\cup Y_{< 0}$ 
\STATE $s\leftarrow [0,0]$; $i\leftarrow 1$; $\pi = \lambda j. j$
\WHILE{$R\not= \emptyset$}
	\IF{$\exists$ $j\in R$ such that $s+\mathcal{S}(\code{y}_j)\subseteq M$}
		\STATE $s\leftarrow s+\mathcal{S}(\code{y}_j)$; $R\leftarrow R\setminus j$
		\STATE $\pi(i)\leftarrow j$; $i\leftarrow i+1$
	\ELSE
		\STATE Fail
	\ENDIF
\ENDWHILE
\FORALL{$j\in Y_{\top}$}
	\STATE $s\leftarrow s+\mathcal{S}(\code{y}_j)$; $\pi(i)\leftarrow j$; $i\leftarrow i+1$
	\IF{$s\not\subseteq M$}
	\STATE Fail
	\ENDIF
\ENDFOR
\end{algorithmic}
\end{algorithm}

\begin{figure}[tb]
\hrule 
\smallskip
 \centerline{\includegraphics[width=0.9\linewidth]{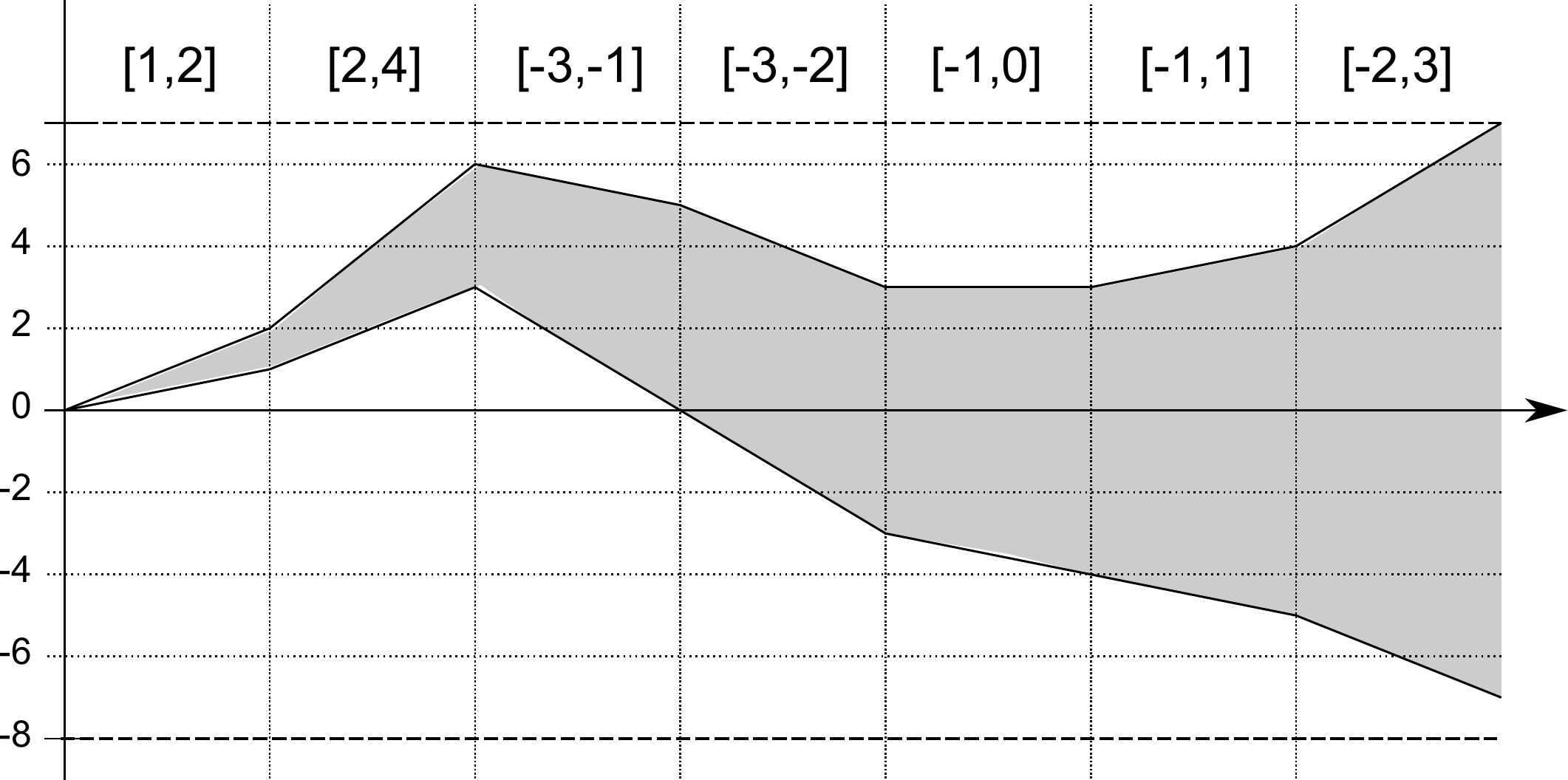}}
\smallskip
\hrule 
 \caption{The graphical evolution of the partial sums of the repaired expression in Ex.~\ref{ex:repaired}.}
  \label{figalgosum}
\end{figure}

\begin{example}
\label{ex:repaired}
Let us consider the Ex.~\ref{ex3}, with $b = 4$.
In this case $M_4=[-8,7]$ and an overflow arises when the term $\code{x}_6 \in [-1,1]$ is added since the partial sum evaluates to
$[-9,3] \not\subseteq M_4$. 
In Algorithm \ref{algosum}, $Y_{>0}= \{ 3,  7\}$, $Y_{<0}= \{ 4, 5 \}$ and $Y_{\top}= \{ 1, 2, 6\}$. 
One output of the algorithm is (the permutation corresponding to) the sum:
$\code{x}_3 + \code{x}_7 + \code{x}_4 + \code{x}_5 + \code{x}_2+ \code{x}_6 + \code{x}_1$.
\qed 
\end{example}

The Fig.~\ref{figalgosum} provides a graphical intuition of the way in which the Algorithm \ref{algosum} works.
It starts with the interval $[0,0]$, and it greedily selects variables such that the current partial sum gets closer, \emph{e.g.}, to $\overline{M}$.
When any value in $Y_{>0}$ causes the partial sum to go above $\overline{M}$, then elements from $Y_{<0}$ are selected. 
Graphically this means that the interval is heading toward $\underline{M}$.
The algorithm continues until $R_0$ is empty, or it is not possible to ``correct'' the partial sum.
An interesting yet unexpected result is the following one, stating that the algorithm is complete, \emph{i.e.}, if there exists a solution to  $\mathbb{R}_{\langle \code{L}, \code{L}, \mathsf{Intv}\rangle}$ then our algorithm finds it.
If it fails, there is no way to repair \code{e} using $\mathcal{S}$.

\begin{property} \textit{(Completeness)}
\label{propsum}
Let $\asem{P}(\code{pp}) = \mathcal{S}$ be an abstract element belonging to $\mathsf{Intv}$.
If Algorithm \ref{algosum} fails then there exists no permutation $\pi \in [1,k]\rightarrow [1,k]$ of the elements of the sum
such that the partial sums of $\sum_{i=1}^{n}\code{y}_{\pi(i)}$ do not overflow, i.e. there is no $\code{e'} \in \mathtt{L}$ such that $\code{e} \equiv_\mathbb{Z} \code{e'}$ 
and $\notOverflow{\gamma(\mathcal{S})}{\code{e'}}$.
\end{property}

\smallskip
\textit{Proof}
First, let us focus on the while loop of Line 5. We show that if Algorithm \ref{algosum} fails at Line 10 then
any parsing of the sum leads to an overflow.
More precisely, if $\exists Y'$ and $Y''$ such that $Y'\cup Y''=[1,k]$, $Y'\cap Y''=\emptyset$, $s=\sum_{i\in Y'} \code{y}_i\subseteq M$ 
and $\forall i\in Y''$, $s+\code{y}_i\not\subseteq M$ then $\sum_{i=1}^k \code{y}_i \not\subseteq M$.

Let $Y''=\{b_1,\ldots , b_m\}$ and let us focus on the upper bound. 
If $\forall i,\  1\le i\le m,\ \overline{s}+\overline{\code{y}}_{b_i}> \overline{M}$ then
\[
\begin{array}{ccc}
\overline{s}+\overline{\code{y}}_{b_1} & > & \overline{M} \\
\overline{s}+\overline{\code{y}}_{b_2} & > & \overline{M} \\
\vdots &  &\vdots\\
\overline{s}+\overline{\code{y}}_{b_m} & > & \overline{M}.
\end{array}
\]
Their sum implies that
\begin{equation}\label{eqcasc}
m\overline{s}+\sum_{b\in Y''} \overline{\code{y}}_b > m\overline{M}.
\end{equation}
Using the fact that $s=\sum_{i\in Y'} \code{y}_i$ and that $Y'\cup Y''=[1,k]$, Equation (\ref{eqcasc}) can be transformed as follows:  
\begin{eqnarray*}
&m\overline{s}+\sum_{b\in Y''} \overline{\code{y}}_b > m\overline{M}\\
\Leftrightarrow & (m-1)\overline{s}+\overline{s}+\sum_{b\in Y''} \overline{\code{y}}_b > m\overline{M}\\
\Leftrightarrow & (m-1)\overline{s}+\sum_{a\in Y'} \overline{\code{y}}_a+\sum_{b\in Y''} \overline{\code{y}}_b > m\overline{M}\\
\Leftrightarrow &
\sum_{i=1}^k \overline{\code{y}}_i > m\overline{M}-(m-1)\overline{s}.
\end{eqnarray*}
Now, let us assume that $\overline{s}\le \overline{M}$. 
From the previous equation it follows that
$$
\sum_{i=1}^k \overline{\code{y}}_i > m\overline{M}-(m-1)\overline{s} \ge m\overline{M}-(m-1)\overline{M}.
$$
Therefore, $\sum_{i=1}^k \overline{\code{y}}_i> \overline{M}$, which  is a contradiction. 
The same arguments holds for the lower bound $\underline{M}$. 

Concerning the for loop of Line 13, any element $\code{y}_i$ with $i\in Y_{\top}$ makes the bounds of $s$ grow (except for the trivial case of adding [0, 0]). 
Consequently, if we cannot fix the variables in $R$, there is no way we can fix the initial set of variables.
The interval sum is deterministic, so the partial sum after the while loop of Line 5 is always the same interval, no matter which parsing is inferred.
In the loop of Line 13, the width of the interval for the partial sum can only grow at each step  with 
one of the two bounds strictly moving toward $\overline{M}$ or $\underline{M}$. As a consequence, if some parsing fails then all the parsings fail.
\qed

\begin{property} \textit{(Complexity)}
For a linear expression with $O(k)$ variables, Algorithm \ref{algosum} performs $O(k^2)$ operations. 
\end{property}

\textit{Proof}
Constants are bounded and, consequently, the preprocessing may only add a constant multiplicative factor.
So the number of terms in the sums given to Algorithm \ref{algosum} is asymptotic linear in the number of variables of the source expressions. 
The while loop of Line 5 is executed
$O(k)$ times and the selection of $j$ at Line 6 is done in $O(k)$. 
The for loop of Line 13 being linear in $k$,  the global complexity of the algorithm is $O(k^2)$.\qed

\begin{algorithm}[tb]
\caption{Boolean Expressions Refactoring}\label{algorel}
\begin{algorithmic}[1]
\REQUIRE Two sets of variables $A$, $B$, a Boolean expression  $\code{b}=\code{e}_1 \Diamond \code{e}_2$ such that $\code{e}_1 = \sum_{\code{a}\in A} \code{a}$ and $\code{e}_2=\sum_{\code{b}\in B} \code{b}$ and an interval map $\mathcal{S}_0$.
\smallskip

\ENSURE A solution to the overflowing problem, or fail if it does not exist
\medskip

\STATE $\cal{M} \leftarrow \emptyset$; $\code{e}'_1 \leftarrow \code{e}_1$; $\code{e}'_2 \leftarrow \code{e}_2$; $\mathcal{S} \leftarrow \mathcal{S}_0$ 
\WHILE{$\neg$Algorithm\ref{algosum}($\code{e}'_1, \mathcal{S}$) $\vee \neg$Algorithm\ref{algosum}($\code{e}'_2, \mathcal{S}$)}
	\IF{$d(\code{e}'_1)\ge d(\code{e}'_2)$}
		\STATE $\alpha \leftarrow \mathsf{select}(\mathcal{S}(A \setminus \cal{M}))$
                \IF{$\alpha$ = \code{fail}}
                \STATE Fail
                \ENDIF
                \STATE $\code{y} \leftarrow \mathcal{S}^{-1}(\alpha)$
                \STATE $A \leftarrow A \setminus \{\code{y}\}; B \leftarrow B \cup \{ \code{y}\}$; 
	\ELSE
		\STATE $\beta \leftarrow \mathsf{select}(\mathcal{S}(B \setminus \cal{M}))$
                \IF{$\beta$ = \code{fail}}
                \STATE Fail
                \ENDIF
                \STATE $\code{y} \leftarrow \mathcal{S}^{-1}(\beta)$
                \STATE $A \leftarrow A \cup \{\code{y}\}; B \leftarrow B \setminus \{ \code{y}\}$
	\ENDIF
        \STATE $\code{e}'_1 \leftarrow \sum_{\code{a} \in A} \code{a}$; $\code{e}'_2 \rightarrow \sum_{\code{b} \in B} \code{b}$  
	\STATE $\cal{M} \leftarrow \cal{M} \cup \{ \code{y} \}$
        \STATE $\mathcal{S} \rightarrow \mathcal{S}[\code{y} \mapsto -\mathcal{S}]$        
\ENDWHILE
\RETURN $\sum_{\code{a} \in A} \phi_{\mathcal{S}_0, \mathcal{S}}(\code{a}) \Diamond \sum_{\code{b} \in B} \phi_{\mathcal{S}_0, \mathcal{S}}(\code{b})$
\end{algorithmic}
\end{algorithm}

\section{Relations among Linear Expressions}
\label{sec:Relations}

We extend the results of the previous section when the considered expressions are comparisons of linear arithmetic expressions with unary coefficients.
We consider the set 
\[
\code{B} = \{ \code{e}_1 \ \Diamond\ \code{e}_2 \mid \code{e}_1, \code{e}_2 \in \code{L},  \Diamond\in \{ =,\ \not=,\ <,\ >,\ \le,\ \ge\}\} 
\]
for input and output expressions.
We use the abstract domain $\mathsf{Intv}$ for the semantic knowledge.
The Algorithm \ref{algorel}  computes an \emph{under}-approximation  to $\mathbb{R}_{\langle \code{B}, \code{B}, \mathsf{Intv}\rangle}$ in \emph{polynomial} time.

We first need to define some functions used by the algorithm.
The (distance from zero) function $d \in \mathbb{Z} \times \mathbb{Z} \rightarrow \mathbb{Z}$  is such that 
\[
d([\underline{x},\overline{x}])=\max\big(|\underline{x}|,|\overline{x}|\big).
\]

We let $d(\code{e})$ denote $d(\mathsf{eval_{Intv}}(\code{e}, \asem{P}(\code{pp}))$, where $\mathsf{eval_{Intv}}$ is the \emph{most} precise evaluation of \code{e} using \emph{unbounded} interval arithmetics~\cite{Intervals}. 

The function  
\[ \mathsf{select}(X) \in \wp(\mathbb{Z} \times \mathbb{Z}) \rightarrow (\mathbb{Z} \times \mathbb{Z}) \cup \{\code{fail}\}
\]
  selects an interval $x \in X$ such that $\underline{M}\not\in x$ and $\forall y\in X. d(y)\le d(x)$.
If such an interval does not exist, then $\mathsf{select}(X) = \code{fail}$. 

Given an (abstract) environment mapping variables to intervals  $\mathcal{S} \in \code{Vars} \rightarrow (\mathbb{Z} \times \mathbb{Z})$, we assume, without losing generality, that $\mathcal{S}$ is injective --- the Algorithm can be easily extended otherwise. 

Finally, we let $\phi_{\mathcal{S}_0, \mathcal{S}} \in \code{Vars} \rightarrow \code{E}$ denote the sign of a variable.
It is  defined as $\phi_{\mathcal{S}_0, \mathcal{S}}(\code{x}) = \code{x}$ if $\mathcal{S}_0(\code{x}) = \mathcal{S}(\code{x})$; $\phi_{\mathcal{S}_0, \mathcal{S}}(\code{x}) = -1 * \code{x}$ if  $\mathcal{S}_0(\code{x}) = -\mathcal{S}(\code{x})$ and undefined otherwise.

The idea of Algorithm~\ref{algorel} is to move terms to the left or the right of the $\Diamond$ relation, and then check whether they overflow.
If it cannot pick any variable, then it simply fails.
The input of the algorithm are two sets of variables, $A$ and $B$, appearing as sum on the left and on the right of $\Diamond$.
The set $\mathcal{M}$ remembers which variables have been moved on the left or right side.
On  entry it is trivially the empty set.
The arithmetic expressions $\code{e}'_1, \code{e}'_2$ are the current approximations for the solution, and $\mathcal{S}$ records the ranges for variables.
On entry, those variables are set to the input parameters.

The Algorithm~\ref{algorel} iterates as long as at least one of the expressions $\code{e}'_1$ or $\code{e}'_2$ overflows --- we use Algorithm~\ref{algosum} to check it. 
It first selects the expression which is further away from zero --- roughly, the expression which causes the largest overflow.
Then, it selects the addendum that contributes most to the overflow, and moves it on the other side of $\Diamond$.
This is reflected by the updates to $A$ and $B$.
The memoization set $\cal{M}$ remembers which terms have already been moved.
This forbids the same term to be moved twice. 
Finally, $\mathcal{S}$ is updated to remember that \code{y} is negated.
The algorithm fails if $\mathsf{select}$ fails, that is if no more term can be moved to the other side of the relation.
Otherwise, a new expression is returned, where variables moved to one side or another of $\Diamond$ are negated.

\begin{example}
Let us  consider the evaluation of the Boolean expression  
\(
\code{x}_1+\code{x}_2\le \code{x}_3+\code{x}_4+\code{x}_5
\)
with simplified 4 bits integers and with the semantic information:
\[
\mathcal{S}_0 = \mathcal{S} = \{  \code{x}_1 \in [-1,1],  \code{x}_2 \in[-2,0], \code{x}_3 \in[1,2], \code{x}_4 \in [2,3], 
 \code{x}_5 \in [5, 6]
\}.
\]  
An overflow arises in the evaluation of the right hand side of the relation since: 
\[\code{x}_3+\code{x}_4+\code{x}_5\in[8,11] \not \subseteq [-8, 7].\] 
Let us apply Algorithm~\ref{algorel}.
At the first step, $B=\{ \code{x}_3,\code{x}_4,\code{x}_5\}$, $\mathcal{M}=\emptyset$. 
Therefore $\mathsf{select}(\mathcal{S}(B\setminus \mathcal{M})) = [5,6]$ since $d(\code{x}_5)=6$. 
The memoization set is updated to  $\mathcal{M}=\{\code{x}_5\}$ and the relation is transformed into
\(
\code{x}_1+\code{x}_2+\code{x}_5 \le \code{x}_3+\code{x}_4.
\)
We record the fact that $\code{x}_5$ appears in a negated context by updating the semantic information:
\[
\mathcal{S} = \{  \code{x}_1 \in [-1,1],  \code{x}_2 \in[-2,0], \code{x}_3 \in[1,2], \code{x}_4 \in [2,3],  \mathbf{\code{x}_5 \in [-6, -5]}\}.
\]
Now, $ \code{x}_1+\code{x}_2+\code{x}_5\in [-9,-4] \not \subseteq[-8, 7]$.
Therefore, the evaluation of the left operand may lead to an overflow, and so the loop is entered again.
The function $\mathsf{select}(\mathcal{S}(A\setminus \mathcal{M}))$ selects $[-2, 0]$, that is the interval corresponding to $\code{x}_2$.
The new expression is
\( \code{x}_1+\code{x}_5 \le \code{x}_3+\code{x}_4+\code{x}_2, \)
with 
\[
\mathcal{S} = \{  \code{x}_1 \in [-1,1],  \mathbf{\code{x}_2 \in[0,-2]}, \code{x}_3 \in[1,2], \code{x}_4 \in [2,3],  {\code{x}_5 \in [-6, -5]}\}.
\]
The algorithm stops since the expression does not overflow.
The expression
\(
 \code{x}_1-\code{x}_5 \le \code{x}_3+\code{x}_4-\code{x}_2
\)
is then returned to the caller.
\qed
\end{example}

\begin{property}
For a relation between two linear expressions with $O(n)$ variables, Algorithm \ref{algorel} performs $O(n^3)$ steps. 
\end{property}

\textit{Proof}
Each variable is moved at most once from one side of the relation to the other. 
Then the while loop of Line 2 is repeated $O(n)$ times.
The $\mathsf{select}$ function is  $O(n)$ but the Algorithm~\ref{algosum} is $O(n^2)$.
The global complexity is then $O(n^3)$.\qed

\section{Exploiting Relational Information}
\label{sec:Octagons}

We extend the results of Sect.~\ref{sec:sum} by using $\mathsf{Oct}$, the abstract domain of Octagons~\cite{Mine-HOSC06} as the underlying semantic knowledge. 
Intuitively, having a more precise abstract domain enables the inference of more relations among variables and as a consequence more arithmetic expressions can be repaired. 
On the other hand, it also makes the search space for the  problem larger.
We show that there exists an algorithm to solve $\mathbb{R}_{\langle \mathcal{I}, \mathcal{O}, \mathsf{Oct}\rangle}$ exactly, \emph{i.e.}, if an expression can be repaired, then our algorithm finds it.

With $\mathsf{Oct}$, constraints are either in the form $a\le \code{x} \le b$ (non-relational constraints) or  $a\le \code{x} \pm \code{y}\le b$ (relational constraints).
In the following  we will use an interval notation to denote those constraints:  $\mathcal{S}(\code{x})=[a,b]$ and $\mathcal{S}(\code{x+y})=[a,b]$.

\begin{example}\label{exoct1}
Let us consider variables \code{x, y, z} of type $M_4$, the sum $\code{x+y+z}$, and the semantic information $\mathcal{S}$:
\[
\{
\code{x}\in [-2,2], \code{y}\in [-1,3], \code{z}\in [-1,4], \code{x+y} \in [-2, 3], \code{y+z} \in [-2,  4],  \code{x+z} \in [-1, 5]
\}.
\]
If we ignore the relational constraints then we cannot propose a fix for the expression. In fact,  Algorithm~\ref{algosum} will fail because: 
\[
[-2,2]+[-1,3]+[-1,4]\not\subseteq [-8,7].
\]
 \qed
\end{example}

We assume the application of the same preprocessing steps of the previous sections (multiplications by constants are expanded into sums, variables are only added -- subtraction is captured in $\mathcal{S}$).
Furthermore, for the sake of simplicity, we also assume that all the variables are different --- variables with multiple occurrences are renamed.

Our repairing algorithm performs a depth-first visit of a weighted directed graph $G$.
Intuitively, the graph indicates in which order we should  add variables or pairs of variables for which we have some relational information in $\mathcal{S}$.
We do not explicitly build the graph --- it is of exponential size.  
Instead, we memorize only the current path corresponding to a non overflowing partial sum.
The nodes correspond to partial sums of variables. 
The weight (an interval) attached to an edge $e$ indicates the  value which must be added to the partial sum when one or two variables are added.
In general, there exists several paths from a source node to a destination node depending on which constraints we use to add the variables. 
For instance, in Ex.~\ref{exoct1}, we have a path corresponding to $\code{x}+(\code{y}+\code{z})$ and another path corresponding
to $(\code{x}+\code{y})+\code{z}$ whose weights are $[-4,6]$ and $[-3,7]$, respectively.

Formally, given an input set of variables $X=\{\code{x}_1,\ldots\code{x}_n\}$ and the semantic information $\mathcal{S}$,  we define the graph $G=(V,E)$ as follows:
\begin{itemize}
\item $V=\wp(X)$: the nodes of $G$ are sets of variables,
\item the entry node is $\emptyset$ and the exit node is $X$,
\item $(v_1,v_2)\in E$ if $v_2=v_1\cup\{ \code{x}_i\}$ for some $v_1\subseteq X$ and some variable $\code{x}_i\in X$.
In this case, $\mathcal{W}\big((v_1,v_2)\big)=\mathcal{S}(\code{x}_i)$,
\item $(v_1,v_2)\in E$ if $v_2=v_1\cup\{ \code{x}_i,\code{x}_j\}$ for some $v_1\subseteq X$ and  some pair of variables $\code{x}_i$ and $\code{x}_j$, with $i \neq j$.
In this case, $\mathcal{W}\big((v_1,v_2)\big)=\mathcal{S}(\code{x}_i+\code{x}_j)$.
\end{itemize}

\begin{figure}[tb]
\hrule 
\smallskip
 \centerline{\includegraphics[width=10cm]{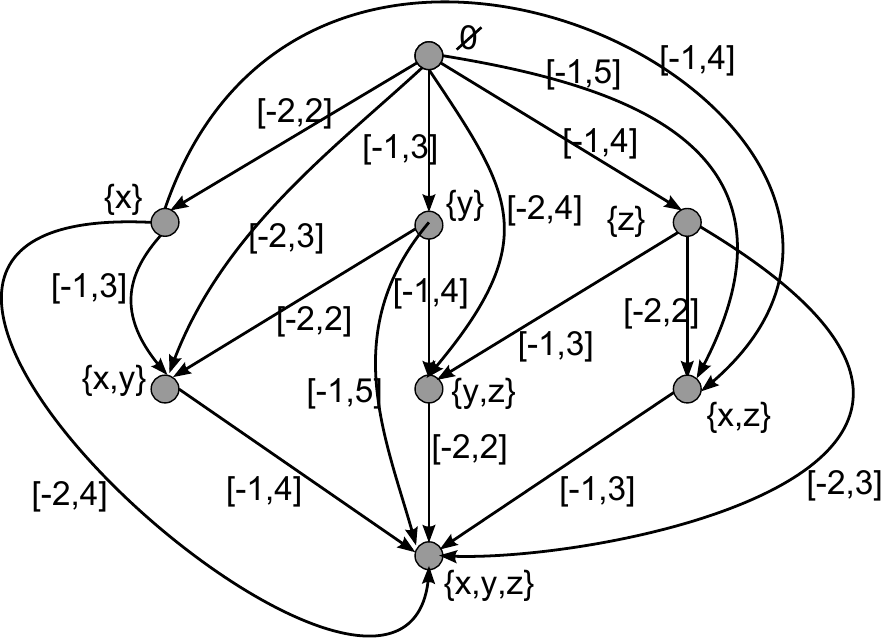}}
\smallskip
\hrule 
 \caption{
Graph describing the ways of summing \code{x}, \code{y}, \code{z} with the information of Ex.~\ref{exoct1}.}
  \label{figalgooct}
\end{figure}

\begin{example}
Let us consider the arithmetic expression and the semantic information of the Ex.~\ref{exoct1}.
The Figure~\ref{figalgooct} shows the corresponding graph $G$.
With $G$, we can derive the non-overflowing expressions $\code{(x+y)+z}$  or $\code{x+(y+z)}$.
The expression $\code{y+(x+z)}$ is not a repair -- it may overflow under the knowledge encoded by $\mathcal{S}$.
\qed
\end{example}

\begin{algorithm}[tb]
\caption{Sum Refactoring with Relational Information}\label{algosumoct}
\begin{algorithmic}[1]
\REQUIRE A path $\overrightarrow{v}$ on $G$ and an octagon $\mathcal{S}$
\smallskip

\ENSURE A path indicating how to compute the total sum without overflows, or fail if it does not exists
\medskip
\STATE{\textbf{Function} $\mathsf{Repair}(\overrightarrow{v})$}
\STATE $\mathcal{N} \leftarrow \mathit{succ}(\overrightarrow{v})$;  $\overrightarrow{r}\leftarrow \overrightarrow{v}$ 
\WHILE{$\mathcal{N}\not= \emptyset \wedge \mathit{dest}(\overrightarrow{r})\not= X$}
	\STATE $v'\leftarrow  \mathit{pickOne}(\mathcal{N})$
	\IF{$\notOverflow{\mathcal{S}}{\sum_{\code{v} \in \overrightarrow{v}\curvearrowright v'} \code{x}}$}
		\STATE $\overrightarrow{r}\leftarrow \mathsf{Repair}(\overrightarrow{v}\curvearrowright v')$
	\ENDIF
	\STATE $\mathcal{N}\leftarrow \mathcal{N}\setminus \{ v'\}$
\ENDWHILE
\IF{$\mathit{dest}(\overrightarrow{r})=X$}
	\RETURN $\overrightarrow{r}$
\ELSE
	\STATE Fail
\ENDIF
\end{algorithmic}
\end{algorithm}

Let $v_1,\ldots v_k$ be a sequence of vertices in $G$.
Then, $\overrightarrow{v}=(v_1,\ldots v_k)$ is a path if $v_1 = \emptyset$ and for all $i$ such that $1\leq i < k$, $(v_{i}, v_{i+1}) \in E$; $\overrightarrow{v}\curvearrowright v'$ is the path extended with vertex $v'$;  $\mathit{dest}(\overrightarrow{v})=v_k$ is the last vertex in a non-empty path;  $succ(v)$ are the successors of vertex $v$, $\mathit{succ}(\overrightarrow{v})=\mathit{succ}(\mathit{dest}(\overrightarrow{v}_k))$, are the successors of a (non-empty) path. 

The function $\mathsf{Repair}$, Algorithm \ref{algosumoct},  performs a depth-first search in $G$.
The input to $\mathsf{Repair}$ is path $\overrightarrow{v}$ in $G$.
Intuitively, $\overrightarrow{v}$ describes a way of summing the variables in $v_k$.
The weight of $\mathcal{W}(\overrightarrow{v})$ is the \emph{unbounded} interval sum of the weights of the edges:
\[
\mathcal{W}(\overrightarrow{x})=\sum_{j=1}^{k-1} \mathcal{W}\big((v_{j},v_{j+1})\big).
\]

Algorithm \ref{algosumoct} visits $G$ without explicitly building it.
The set  $\mathcal{N}$ is initialized to the set of successor nodes of the current path. 
The algorithm iterates until $\mathcal{N}$ is empty (in which case the algorithm fails) or the exit node $X$ is  reached (in which case the algorithm succeeds).
In the main loop, a new node from the unexplored successors $\mathcal{N}$ is picked via the function $\mathsf{pickOne}$.
Heuristically, we want $\mathsf{pickOne}$ to return the node $v'$ minimizing $\mathcal{W}\big((x_n,x')\big)$. 
If the new partial sum corresponding to the path $\overrightarrow{v}\curvearrowright v'$ does not overflow, then the exploration continues from such a path.
Otherwise, the exploration continues with another successor in $\mathcal{N}$.

\begin{example}
Let us apply Algorithm \ref{algosumoct} to Ex.~\ref{exoct1}. 
The initial value for $\mathsf{Repair}$ is $\emptyset$, \emph{i.e.} the path corresponding to the entry node. 
We have 
\[
\mathcal{N}=\big\{ \{ \code{x}\}, \{ \code{y}\}, \{ \code{z}\},\{ \code{x,y}\}, \{ \code{x,z}\},\{ \code{y,z}\}\big\}.
\]
The function $\mathsf{pickOne}$ selects  $\{\code{x}\}$ since $d((\emptyset,\{\code{x}\}))=2$ is minimal. 
Therefore, it invokes $\mathsf{Repair}$  on the path $(\emptyset,\{ \code{x}\})$. 
It choses  $\{\code{x}, \code{y}\}$ as successor.
At the next recursive call, when it adds $\code{z}$ to the path, it detects an  overflow  since 
\[
[-2,2]+[-1,3]+[-1,4]\not\subseteq [-8,7].
\] 
The algorithm backtracks and finds the path 
\[
(\emptyset,\{\code{x}\},\{\code{x}, \code{y}, \code{z}\})\] whose weight is $[-2,2]+[-2,4]\subseteq [-8,7]$. \qed
\end{example}

\begin{property} \textit{(Completeness)}
Algorithm \ref{algosumoct} is complete. 
\end{property}

\textit{Proof}
Algorithm \ref{algosumoct} performs a search into a graph $G$ which describes all the ways of summing the variables of
the expression. Then, if a solution exists, it is included in $G$ and the algorithm necessarily finds it. \qed

The Algorithm \ref{algosumoct} can be generalized to any template-based numerical abstract domain~\cite{SankaranarayananEtAl-VMCAI05}. 
In this case, we have additional constraints of form 
\begin{equation}\label{eqlinconstr}
a \le a_1\cdot \code{x}_1+\ldots+a_n\cdot \code{x}_n \le b
\end{equation}
where the $\code{x}_i$, $1\le i\le n$ are variables of the program and
where $a$, $b$ and the coefficients $a_i$, $1\le i\le n$ are constants.
The preprocessing transforms each term $a_i\cdot x_i$ into $\code{x}_i^{(1)}+\code{x}_i^{(2)}+\ldots+\code{x}_i^{(a_i)}$ and
the graph used by Algorithm \ref{algosumoct} can be extended with new edges $(v_1,v_2)$ such that
$$
v_2=v_1\cup \bigcup_{1\le i \le n} \left(\bigcup_{1\le j\le a_i}\code{x}_i^{(j)}\right)
$$
for some $v_1\subseteq X$ and  for the constraint of Equation (\ref{eqlinconstr}).
In this case, the weight associated to the new edge $(v_1,v_2)$ is $\mathcal{S}(\code{x}_1+\ldots+a_n\cdot \code{x}_n)$.

\section{Experimental Results}
\label{sec:experiments}

We have a prototype OCAML implementation that we used to validate and to experiment with the algorithms presented in the previous sections. 
We use $32$ bits signed integers, \emph{i.e.}, $M_{32} \in [-2^{31}, 2^{31}-1]$. 

For the algorithms \ref{algosum} and \ref{algorel}, we use the variables $\code{x}_1,\ldots,\code{x}_6$ with
\[
\mathcal{S} = \{\code{x}_1 \in [0, 2^{29}],\code{x}_2 \in [-2^{29}, 0], \code{x}_3 \in [-2^{29}, 2^{29}], \code{x}_4 \in [1,1], \code{x}_5 \in [-1,-1], \code{x}_6 \in [-1,1]\}.
\]
There is nothing special about $\mathcal{S}$, we generated it randomly.

To evaluate Algorithm~\ref{algosum}, we consider \emph{all} the expressions in the  form of
\(
\code{e}=\sum_{i=1}^6 a_i\cdot \code{x}_i
\)
and  $a_i\in \{ 0,1,2,4\}$ for $1\le i\le 6$.
Overall, there  are $3402$ of such expressions. 
Under  $\mathcal{S}$, $1093$ expressions do not overflow, $2268$ cannot be repaired, and  $43$ can be repaired.

Algorithm~\ref{algosum} proves it in a very little time:
The \emph{overall} time to analyze, suggest a repair or prove that such repair does not exists is of $120\mathit{ms}$.  
We report some of the repairs discovered by the algorithm in Fig.~\ref{xpsum}.

\begin{figure}[tb]
\centerline{
\begin{tabular}{c}
\hline
\\
\text{Algorithm 1}\\
$2\cdot\code{x}_1+2\cdot\code{x}_3+2\cdot\code{x}_5+\code{x}_6 \longrightarrow 2\cdot\code{x}_5+2\cdot\code{x}_1+\code{x}_6+2\cdot\code{x}_3$\\
$2\cdot\code{x}_1+2\cdot\code{x}_3+\code{x}_4+2\cdot\code{x}_5 \longrightarrow 2\cdot+\code{x}_5+\code{x}_4+2\cdot\code{x}_1+2\cdot\code{x}_3$\\
$4\cdot\code{x}_1+2\cdot\code{x}_2+\code{x}_4+2\cdot\code{x}_5 \longrightarrow 2\cdot+\code{x}_5+\code{x}_4+2\cdot\code{x}_2+4\cdot\code{x}_1$\\
$4\cdot\code{x}_1+2\cdot\code{x}_2+\code{x}_4+4\cdot\code{x}_5+\code{x}_6 \longrightarrow 4\cdot\code{x}_5+\code{x}_4+2\cdot\code{x}_2+4\cdot\code{x}_1+\code{x}_6$\\
$4\cdot\code{x}_1+2\cdot\code{x}_2+2\cdot\code{x}_4+4\cdot\code{x}_5 \longrightarrow 4\cdot\code{x}_5+2\cdot\code{x}_4+2\cdot\code{x}_2+4\cdot\code{x}_1$\\
$2\cdot\code{x}_1+\code{x}_2+2\cdot\code{x}_3+2\cdot\code{x}_4+4\cdot\code{x}_5+\code{x}_6 \longrightarrow 4\cdot\code{x}_5+2\cdot\code{x}_4+\code{x}_2+2\cdot\code{x}_1+\code{x}_6+2\cdot\code{x}_3$\\
\\
\text{Algorithm 2}\\
$\code{x}_4+\code{x}_5 < 4\cdot \code{x}_1+4\cdot \code{x}_2+4\cdot \code{x}_3 \longrightarrow 4\cdot \code{x}_3+4\cdot \code{x}_1+\code{x}_4+\code{x}_5 < 4\cdot \code{x}_2$ \\
$4\cdot \code{x}_3+4\cdot \code{x}_5 < 4\cdot \code{x}_1+4\cdot \code{x}_2+4\cdot \code{x}_4 \longrightarrow 3\cdot \code{x}_3+4\cdot \code{x}_5 < \code{x}_3+4\cdot \code{x}_1+4\cdot \code{x}_2+4\cdot \code{x}_4$ \\
$2\cdot \code{x}_3+4\cdot \code{x}_5 < 4\cdot \code{x}_1+2\cdot \code{x}_2+4\cdot \code{x}_4 \longrightarrow \code{x}_1+3\cdot \code{x}_4+2\cdot \code{x}_3+4\cdot \code{x}_5 < 3\cdot \code{x}_1+2\cdot \code{x}_2$ \\
$\code{x}_4 < 4\cdot \code{x}_1+\code{x}_2+2\cdot \code{x}_3+4\cdot \code{x}_5 \longrightarrow 2\cdot \code{x}_3+\code{x}_2+\code{x}_4 < 4\cdot \code{x}_5+4\cdot \code{x}_1$ \\
$2\cdot \code{x}_1+4\cdot \code{x}_3+\code{x}_4 < 4\cdot \code{x}_2+4\cdot \code{x}_5 \longrightarrow 3\cdot \code{x}_3 < 2\cdot \code{x}_1+\code{x}_4+\code{x}_3+4\cdot \code{x}_2+4\cdot \code{x}_5$ \\
$4\cdot \code{x}_2+2\cdot \code{x}_4+\code{x}_5 < \code{x}_1+4\cdot \code{x}_3 \longrightarrow \code{x}_3+\code{x}_1+4\cdot \code{x}_2+2\cdot \code{x}_4+\code{x}_5 < 3\cdot \code{x}_3$ \\
\\
\text{Algorithm 3}\\
$2\cdot \code{x}_2+2\cdot \code{x}_3+\code{x}_6 \longrightarrow \code{x}_6+(\code{x}_2+\code{x}_3)+\code{x}_3+\code{x}_2$\\
$2\cdot \code{x}_1+2\cdot \code{x}_3+\code{x}_5+\code{x}_6 \longrightarrow \code{x}_5+\code{x}_6+\code{x}_3+(\code{x}_1+\code{x}_3)+\code{x}_1$\\
$2\cdot \code{x}_1+2\cdot \code{x}_2+2\cdot \code{x}_3+\code{x}_4+\code{x}_5 \longrightarrow \code{x}_4+\code{x}_5+\code{x}_2+(\code{x}_1+\code{x}_3)+\code{x}_3+\code{x}_1+\code{x}_2$\\
\\
\hline
\end{tabular}}
\caption{Some of the repairs  produced by the implementation of our algorithms. For Algorithm~\ref{algosumoct}, more repairs are enabled because of a more refined semantic knowledge. }
\label{xpsum} 
\end{figure}

In order to evaluate Algorithm~\ref{algorel}, we consider all the inequalities in the form of
\(
\sum_{i=1}^5 a_i\cdot \code{x}_i < \sum_{i=1}^5 a_i\cdot \code{x}_i
\)
such that: 
\begin{itemize}
\item[(i)] $a_i\in\{0,1,2,4\}$, for $1\le i\le 5$,
\item[(ii)] a given variable $\code{x}_i$ appears at most in one side of the equation,
\item[(iii)] each side of the equation owns at least one term with coefficient $a_i\not= 0$. 
\end{itemize}

There exist $7290$ of such expressions, and only $213$ are not repaired by Algorithm \ref{algorel}.
The \emph{overall} time to explore all such expressions and to suggest a repair, if any, is of $202\mathit{ms}$.
Example of repairs  are in Figure \ref{xpsum}.

To evaluate Algorithm~\ref{algosumoct}, we add to $\mathcal{S}$ the relational information:
\[
\mathcal{S}(\code{x}_1+\code{x}_3)=\mathcal{S}(\code{x}_1+\code{x}_2)=[-2^{29},2^{29}].
\]
With this information, we can repair some more expressions than with Algorithm~\ref{algosum} --- with a negligible additional cost.
Such cases are in Fig.~\ref{xpsum}.

Overall, in our randomly generated tests, the Algorithms seem to perform extremely well.

\section{Related Work}
\vspace{-0.2cm}
The dynamic analysis community has been interested for a while to the problem of repairing faulty programs, \emph{i.e.} programs failing some test cases~\cite{PezzeEtAl11}.
Their approaches can be abstracted as follows.
When a bug is found in a program~\code{P}, a modified program~\code{P'} is generated such that \code{P'} behaves as \code{P} on a certain set of tests but it does not manifest the bug.
The repaired program \code{P'} is generated searching a program in the set of program transformations/mutations of \code{P} --\emph{e.g.}, applying a set of templates~\cite{PerkinsEtAl09,WeimerEtAl09,LeGouesEtAl-ICSE12,SaminiEtAl-ICSE12}.
This approach has many drawbacks.
First, it requires running the program, first to find the bug in~\code{P} and then to generate and check the candidates for \code{P'}. 
This may not be practical because the test suite can be  very large -- in real software up to several hours.
As a consequence the search for \code{P'} is \emph{de facto} impossible for a large \code{P}.
Second, the whole program may not be available -- this is the case for instance when developing libraries.
Third, the repair is as good as the test suite itself: \emph{e.g.}, how can we be sure that an arithmetic overflow has been removed for all the possible inputs and not just for the particular test case?
Fourth, the generated repairs are often counter-intuitive, and not realistic, in that they perturb the semantics of the original program in the ``good'' runs~\cite{WeiEtAl}.
We differentiate from those approaches because our approach is completely static, it exploits the semantic information inferred by the static analyzer, and the repair is guaranteed to only improve the good executions, while removing overflows (``bad'' executions).

Orthogonally, a recent paper from Coker and Hafiz~\cite{CokerHafiz-ICSE13}, introduces a set of refactorings to fix overflows in C programs to be applied in a IDE.
However, unlike us, they do not propose any way to automatically generate such fixes.

The bases for our paper are~\cite{LogozzoBall-OOPSLA12} and~\cite{Martel07,Martel09}. 
In~\cite{LogozzoBall-OOPSLA12} we  introduced the concept of verified repair, the idea is that one can generate a program repair from the semantic information inferred by the static analyzer, and verify that it  improves the program by removing bad behaviors while increasing good ones, up to some level of abstraction.
We proposed some fixes for common bugs, including a greedy incomplete algorithm to fix overflowing expressions.
Here we improve over such an algorithm, by first making clear the concept of input and output language, and that  of the underlying semantic information used to repair the overflowing expression.
Furthermore, in the present paper we present three algorithms for repairing expressions in common cases, and we formally study their properties (complexity and completeness).
In~\cite{Martel07,Martel09} we considered the problem of generating (an approximation of) the most precise arithmetic expression over floating-point values \cite{MullerEtAl2010}.
The problem of improving a floating-point expression presents a slightly different challenge than the one considered here in that, in general, the expressions cannot
be entirely repaired. Because the floating-point arithmetic introduces roundoff errors, it is in general impossible to find an equivalent expression which computes
the exact result, \emph{i.e.}, the result that we would obtain with real numbers. 
A verified repair then is an expression which is equivalent to the original one when interpreted
over the real numbers and which is more accurate than the original expression in the sense that it returns a value closer to the mathematical result when evaluated with floating-point numbers.

\section{Conclusions}
It is our firm belief that  we need to bring static analysis tools to the next level, to make them more practical.
 They should  not limit themselves to find (or prove the absence of) bugs, but they should actively  help the programmers 
 by providing suggestions to improve their program, by removing the bug or by proposing better design choices.

In this paper, we considered the particular yet relevant problem of repairing integer expressions starting from the warnings and the invariants inferred by an abstract interpretation-based static analyzer.
We characterized the three elements in the problem: (i) the input language for expressions (``which expression should I repair?''); (ii) the output language for the expressions (``how am I allowed to repair an expression?''); and (iii) the semantic information (``what do I know about the values of the expression?'').
Then we focused our attention on three common cases, the repairing of: (i) linear expressions with intervals; (ii) Boolean expressions containing linear ones with intervals; and (iii) linear expressions with templates.
We showed that in the first case, quite surprisingly, there exists a complete polynomial algorithm to solve the problem, whereas in the second we have a polynomial yet incomplete one, and in the third we have a complete but worst-case exponential one.

\bibliographystyle{eptcs}
\bibliography{bibi2}

\end{document}